%% file: main.tex
\documentclass[sigconf]{acmart}
\usepackage{graphicx}
\usepackage{tikz,graphics,color,fullpage,float,epsf,caption,subcaption}
\usepackage{pgfplots}
\usepackage{wrapfig}
\setcopyright{none} 
\acmConference[Submission to ACM WPES 2022]{ACM Conference on Computer and Communications Security}{Due 16 July 2022}{Los Angeles}
\acmYear{2022}

\author{Ethan Witwer, James Holland, Nicholas Hopper}
\email{ {witwe004,holla556,hoppernj} @umn.edu}
\affiliation{%
  \institution{University of Minnesota}
}

\begin{document}
\title{Padding-only defenses add delay in Tor} 

\begin{abstract}
\input{00-abstract}
\end{abstract}




\maketitle

\section{Introduction}
\input{01-intro}

\section{Background}
\input{02-background}

\section{Experiments and Results}
\input{03-experiments}

\section{Conclusions and Future Work}
\input{04-conclusion}

\subsubsection*{Acknowledgements}
This work was supported by NSF grant 1815757 and the University of Minnesota Undergraduate Research Opportunities Program (UROP).  

\bibliographystyle{ACM-Reference-Format}
\bibliography{padding-refs}

\end{document}

%% file: 00-abstract.tex
Website fingerprinting is an attack that uses size and timing characteristics of encrypted downloads to identify targeted websites. Since this can defeat the privacy goals of anonymity networks such as Tor, many algorithms to defend against this attack in Tor have been proposed in the literature.  These algorithms typically consist of some combination of the injection of dummy ``padding'' packets with the delay of actual packets to disrupt timing patterns. For usability reasons, Tor is intended to provide low latency; as such, many authors focus on padding-only defenses in the belief that they are ``zero-delay.'' We demonstrate through Shadow simulations that by increasing queue lengths, padding-only defenses add delay when deployed network-wide, so they should not be considered ``zero-delay.'' We further argue that future defenses should also be evaluated using network-wide deployment simulations.

%% file: 01-intro.tex
Tor~\cite{Dingledine2004} is a low-latency anonymity network and web browser used daily by millions of users to evade state and corporate surveillance and censorship. Tor redirects connections through multiple intermediaries using layered encryption so that no single entity knows both the source and destination of a connection.  This hides which sites a user visits and any contents requested from those sites.

However, Tor has been shown to be vulnerable to several traffic analysis attacks, including {\em Website Fingerprinting} (WF) attacks~\cite{hintz2002fingerprinting}. WF attacks use information about the timing, sequence, and volume of packets sent between a client and the Tor network to detect whether a client is downloading a targeted, or ``monitored,'' website.  These attacks have been found to be highly effective against Tor, identifying targeted websites with accuracy as high as 99\%~\cite{bhat2019var} under some conditions.

This has led to many proposed defenses that modify the characteristics of a connection in order to confuse a WF attacker.  A canonical ``high-overhead'' WF defense is BuFlo~\cite{dyer2012peek}: BuFlo sets a constant traffic rate for every connection; in every fixed-length time slot, a Tor node sends one packet on each connection.  If multiple packets are pending on a connection, some must be {\em delayed} to the next slot, and if no packets are pending, an encrypted {\em padding} packet is sent instead. By making all downloads have roughly the same sequence, timing, and volume characteristics, BuFlo greatly reduces the accuracy of WF attacks, but it has a very high cost in terms of the additional bandwidth overhead and latency incurred.

Because of the resource constraints of the Tor network and its focus on providing low latency, the literature contains many proposals for WF defense schemes that have lower costs than BuFlo; we briefly describe several of these in Section~2.  Typically, the cost of these defenses is measured either by {\em trace simulation}, in which the traces of several page downloads over Tor are captured and a simulator is used to add or delay packets in the trace according to the defense; or by implementing the defense as a {\em pluggable transport}~\cite{tor-PT} that encapsulates the connection between a single Tor client and relay in a defended connection.  Researchers then compute the bandwidth overhead by comparing the number of bytes transmitted when downloading a site with and without the WF defense, and if the defense involves adding packet delays, the latency overhead is computed by comparing the overall time to download a site with and without the defense.  

For padding-only defenses, this evaluation will not add any latency, and so such defenses are sometimes referred to as ``zero-delay.''  However, if such defenses were to be implemented on a network-wide scale, the added packets sent as padding would necessarily consume resources that could otherwise be used to send non-padding packets to other clients.  If enough padding is added, such a defense may actually delay connections {\em more} than a defense that uses less padding but sometimes delays packets.  Thus, we contend that it is important when evaluating WF defenses to account for the effects of deploying to the full Tor network.

In this paper, we use the Shadow~\cite{jansen2012shadow} network simulator to evaluate the effect of deploying three padding-only defenses -- REB~\cite{mathews2018understanding}, Spring, and Interspace~\cite{pulls2020towards} -- on a network-wide basis.  By measuring the progress of downloads over time compared to results from the same network without padding, and comparing the download time for files to the number of padding packets injected per download, we show that padding-only defenses cause delay and that more padding causes additional delay.  This illustrates that previous methodologies for measuring delay overhead give an incomplete picture of the costs of WF defense techniques.

\noindent{\bf Our contributions.}  We make the following contributions:
\begin{itemize}
\item We give the first full-network simulations of padding defense deployments.  These show the importance of evaluating additional defenses in this setting.
\item We propose a new methodology for WF defense evaluation, time-to-nth-byte.  Tracking download progress over time gives a more accurate and complete depiction of the delay incurred by a WF defense mechanism.
\item We show that ``zero-delay'' padding defenses cause delay and should not be over-prioritized in comparison with timing-based defenses.
\end{itemize}

%% file: 02-background.tex
\paragraph{Tor.}  The Tor network~\cite{Dingledine2004} is made up of several thousand volunteer-operated {\em relays} distributed globally, which provide service to roughly one million concurrent users at any time~\cite{jansen2016safely}.  To connect to a website using Tor, a client constructs a three-hop circuit consisting of a {\em guard} relay, a {\em middle} relay, and an {\em exit} relay.  The circuit uses layered encryption so that the guard can only see that it forwards from the client to the middle relay, the middle relay can only see that it forwards from a guard to an exit relay, and the exit relay can see the destination traffic but not the client identity.  All traffic between the client and relays is encapsulated into 512-byte {\em cells}.

\paragraph{Website fingerprinting attacks.} 
While neither an ISP on the network path between the client and the guard nor the guard relay itself can see the {\em contents} or ultimate {\em destination} of Tor cells, they can observe the timing, direction, and volume of cells sent in each direction on this connection, which are fairly consistent for any given website. WF attacks use statistical and machine-learning techniques to build ``fingerprints'' of these sequences for a set of targeted websites. A series of results~\cite{Herrmann2009,Lu2010,Wang2013,Hayes2016,Panchenko2016,rimmer2018automated,sirinam2018deep,bhat2019var,rahman2019tik} has shown with increasing accuracy that with no defenses in place to modify these sequences, WF attacks can identify visits to a targeted web page with over 99\% accuracy in some settings.

\paragraph{Defenses.}  To defend against such attacks, it is necessary to modify the sequence of cells observed by the attacker.  One way to do this is to inject extra ``dummy cells'' into either the connection between the client and the guard relay or the tunnel between the client and the middle relay (depending on whether the guard is considered a potential WF attacker or not).  Based on the results of Shmatikov and Wang~\cite{shmatikov2006timing}, and Juarez {\em et al.}~\cite{juarez2016toward}, the Tor project has implemented a circuit padding framework~\cite{tor-circuitpadding} that can deploy stochastic state machines that adaptively pad a connection to fill unlikely gaps between cells.  Matthews, Sirinam, and Wright used this framework to implement the ``Random Extend Burst'' (REB) defense mechanism~\cite{mathews2018understanding}, and Pulls used genetic algorithms to find the locally optimal padding machines Spring and Interspace~\cite{pulls2020towards}.  Several other defenses~\cite{BiMorph2019,DFD2020,gong2020zero} also use padding to disrupt fingerprints, but none have been implemented in the circuit padding framework.

In addition to padding, a defense could elect to delay the transmission of some actual cells to disrupt fingerprints, in an attempt to make a website's fingerprint match a different sequence, such as constant-rate traffic~\cite{cai2012touching,cai2014systematic}, A ``decoy trace''~\cite{wang2017walkie}, a common but evolving traffic rate~\cite{lu2018dynaflow,holland2022regulator}, or an adversarially generated pattern~\cite{BANP21,gong2022surakav}.  These defenses often have more easily-stated security arguments, but because they induce latency by sometimes delaying cells, they are seen as unacceptable in the low-latency context of Tor.

\paragraph{Attack and Defense Evaluation}
Juarez {\em et al} \cite{juarez2014critical} argued in 2014 that research on WF attacks was unfairly privileging the {\em attacker} by restricting to an unrealistic setting. These advantages include the use of ``closed world'' assumption in which only a fixed number of websites could be visited, the ``single tab browsing'' assumption, the use of website frontpages rather than subpages, and the lack of consideration of the effects of network conditions and retraining.  Subsequent works have relaxed many of these conditions, though recently Cherubin, Jansen, and Troncoso~\cite{cherubin2022online} showed that modern techniques still degrade quickly with the size of the ``monitored'' set.

In contrast, Wang~\cite{wang2021one} and Pulls and Dahlberg~\cite{pulls2020website} have argued that WF {\em defenses} should be evaluated in the most optimistic setting for attacks, assuming a single monitored page and even an oracle that can eliminate false positives, because preventing these attacks also protects against weaker attacks.

Since few defense mechanisms can provide such strong protection, we advocate for a more realistic evaluation of the {\em deployment cost} of WF defenses.

%% file: 03-experiments.tex
\subsection{Experimental Setup}

The main tool we used to collect data was Shadow, a network simulator designed to enable realistic and reproducible experiments with Tor. Shadow allows configuration of a network {\em graph}, which specifies nodes (network-connected hosts), how they are connected, and the processes that they should run. We sought to create a network graph resembling the live Tor network, with relays running the Tor process, and clients and servers running Tor and other processes to send and receive traffic over Tor.

We used TorNetTools~\cite{neverenough-sec2021} for this purpose. TorNetTools uses Tor Project metrics~\cite{wecsr10measuring-tor} to create {\em models} of the Tor network, which can be scaled down to account for resource constraints, specifically memory limitations, while remaining representative of the composition of the network. However, this scaling process involves random sampling, which can introduce error that might affect the conclusions of an experiment. Similarly, each simulation of a model involves random sampling, which could also introduce error.

To account for this, we generated multiple models and ran several simulations on each to produce our results. Using Tor Project metrics from March 2022, we generated two models at 0.5\% of the scale of the live Tor network at that time and two models at 0.75\% scale. We ran 5 simulations of each model with the default Tor configuration, which does not include any WF defenses, for a total of 20 simulations. We also compiled Tor with the padding-only defenses REB~\cite{mathews2018understanding}, Spring, and Interspace~\cite{pulls2020towards}, running a total of 20 simulations with each defense.

The models generated by TorNetTools consist of Tor relays; clients and servers that generate realistic network traffic and send it over Tor; and 100 {\em benchmarking clients}, which repeatedly perform 50 KiB, 1 MiB, and 5 MiB downloads throughout a simulation. We used these benchmarking clients to measure bandwidth overhead, delay, and failure rate. We begin by looking at bandwidth overhead, the typical measurement used when reporting the cost of a padding-only defense, with standard Tor, REB, Spring, Interspace.

\begin{figure*}[t]
\centering
\begin{subfigure}[b!]{0.31\textwidth}
    \begin{tikzpicture}
      \begin{axis}[
          width=\linewidth, 
          grid=major, 
          grid style={dashed,gray!30}, 
          xlabel=1MiB Download progress (KiB),
          ylabel=Bandwidth overhead (\%),
          legend pos=north west,
          no markers,
          every axis plot/.append style={very thick},
          scaled x ticks=false,
          scaled y ticks=false
        ]
        \addplot
        table[x=KiB_count,y=CTL_bwoh,col sep=comma] {PCTB_1M.csv};
        \addplot
        table[x=KiB_count,y=SPR_bwoh,col sep=comma] {PCTB_1M.csv}; 
        \addplot
        table[x=KiB_count,y=INT_bwoh,col sep=comma] {PCTB_1M.csv};
        \legend{Control,Spring,Interspace,REB}
      \end{axis}
    \end{tikzpicture}
\end{subfigure}
\hspace{1cm}
\begin{subfigure}[b!]{0.31\textwidth}
    \begin{tikzpicture}
      \begin{axis}[
          width=\linewidth, 
          grid=major, 
          grid style={dashed,gray!30}, 
          xlabel=5MiB Download progress (KiB),
          ylabel=Bandwidth overhead (\%),
          legend pos=south east,
          no markers,
          every axis plot/.append style={very thick},
          scaled y ticks=false
        ]
        \addplot
        table[x=KiB_count,y=CTL_bwoh,col sep=comma] {PCTB_5M.csv};
        \addplot
        table[x=KiB_count,y=SPR_bwoh,col sep=comma] {PCTB_5M.csv};
        \addplot
        table[x=KiB_count,y=INT_bwoh,col sep=comma] {PCTB_5M.csv};
      \end{axis}
    \end{tikzpicture}
\end{subfigure}

\hspace{0.0cm}

\vspace{-12pt}
\caption{{\bf Bandwidth overhead:} Median of receive bandwidth overhead with standard Tor, Spring, and Interspace}
\label{fig-rbw}
\end{figure*}
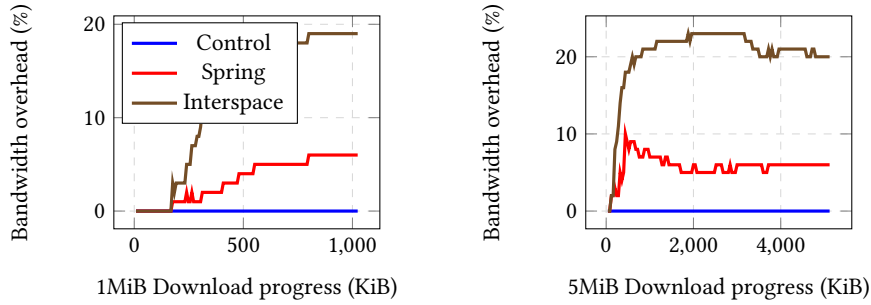
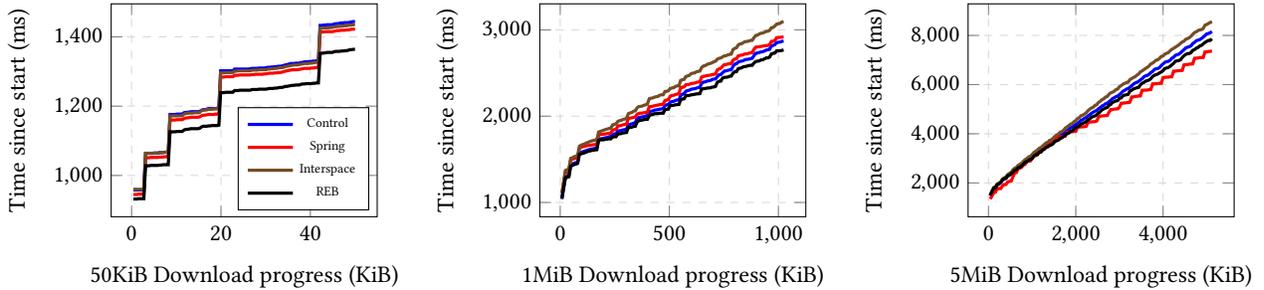
\begin{figure*}
\centering
\begin{subfigure}[b!]{0.31\textwidth}
    \begin{tikzpicture}
      \begin{axis}[
          width=\linewidth, 
          grid=major, 
          grid style={dashed,gray!30}, 
          xlabel=50KiB Download progress (KiB),
          ylabel=Time since start (ms),
          legend pos=south east,
          legend style={font=\tiny},
          no markers,
          every axis plot/.append style={very thick},
          scaled x ticks=false,
          scaled y ticks=false
        ]
        \addplot
        table[x=KiB_count,y=CTL_time,col sep=comma] {TTB_50K.csv};
        \addplot
        table[x=KiB_count,y=SPR_time,col sep=comma] {TTB_50K.csv}; 
        \addplot
        table[x=KiB_count,y=INT_time,col sep=comma] {TTB_50K.csv};
        \addplot
        table[x=KiB_count,y=REB_time,col sep=comma] {TTB_50K.csv};
        \legend{Control,Spring,Interspace,REB}
      \end{axis}
    \end{tikzpicture}
\end{subfigure}
\hfill
\begin{subfigure}[b!]{0.31\textwidth}
    \begin{tikzpicture}
      \begin{axis}[
          width=\linewidth, 
          grid=major, 
          grid style={dashed,gray!30}, 
          xlabel=1MiB Download progress (KiB),
          ylabel=Time since start (ms),
          legend pos=south east,
          no markers,
          every axis plot/.append style={very thick},
          scaled y ticks=false
        ]
        \addplot
        table[x=KiB_count,y=CTL_time,col sep=comma] {TTB_1M.csv};
        \addplot
        table[x=KiB_count,y=SPR_time,col sep=comma] {TTB_1M.csv};
        \addplot
        table[x=KiB_count,y=INT_time,col sep=comma] {TTB_1M.csv};
        \addplot
        table[x=KiB_count,y=REB_time,col sep=comma] {TTB_1M.csv};
      \end{axis}
    \end{tikzpicture}
\end{subfigure}
\hfill
\begin{subfigure}[b!]{0.31\textwidth}
    \begin{tikzpicture}
      \begin{axis}[
          width=\linewidth, 
          grid=major, 
          grid style={dashed,gray!30}, 
          xlabel=5MiB Download progress (KiB),
          ylabel=Time since start (ms),
          legend pos=south east,
          no markers,
          every axis plot/.append style={very thick},
          scaled y ticks=false
        ]
        \addplot
        table[x=KiB_count,y=CTL_time,col sep=comma] {TTB_5M.csv};
        \addplot
        table[x=KiB_count,y=SPR_time,col sep=comma] {TTB_5M.csv};
        \addplot
        table[x=KiB_count,y=INT_time,col sep=comma] {TTB_5M.csv};
        \addplot
        table[x=KiB_count,y=REB_time,col sep=comma] {TTB_5M.csv};
      \end{axis}
    \end{tikzpicture}
\end{subfigure}

\vspace{-12pt}


\caption{{\bf Delay:} Time-to-byte data with standard Tor, Spring, Interspace, and REB
}
\label{fig-ttb}
\end{figure*}
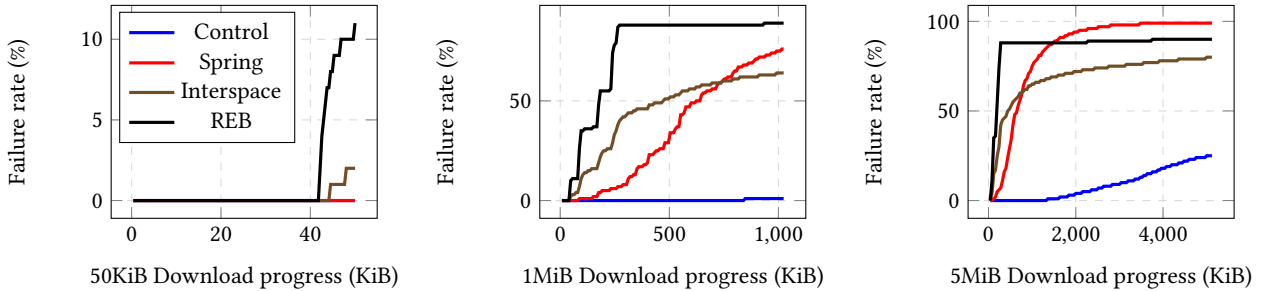
\begin{figure*}[t!]
\centering
\begin{subfigure}[b!]{0.31\textwidth}
    \begin{tikzpicture}
      \begin{axis}[
          width=\linewidth, 
          grid=major, 
          grid style={dashed,gray!30}, 
          xlabel=50KiB Download progress (KiB),
          ylabel=Failure rate (\%),
          legend pos=north west,
          no markers,
          every axis plot/.append style={very thick},
          scaled x ticks=false,
          scaled y ticks=false
        ]
        \addplot
        table[x=KiB_count,y=CTL_error_rate,col sep=comma] {ERR_50K.csv};
        \addplot
        table[x=KiB_count,y=SPR_error_rate,col sep=comma] {ERR_50K.csv}; 
        \addplot
        table[x=KiB_count,y=INT_error_rate,col sep=comma] {ERR_50K.csv};
        \addplot
        table[x=KiB_count,y=REB_error_rate,col sep=comma] {ERR_50K.csv};
        \legend{Control,Spring,Interspace,REB}
      \end{axis}
    \end{tikzpicture}
\end{subfigure}
\hfill
\begin{subfigure}[b!]{0.31\textwidth}
    \begin{tikzpicture}
      \begin{axis}[
          width=\linewidth, 
          grid=major, 
          grid style={dashed,gray!30}, 
          xlabel=1MiB Download progress (KiB),
          ylabel=Failure rate (\%),
          no markers,
          every axis plot/.append style={very thick},
          scaled x ticks=false,
          scaled y ticks=false
        ]
        \addplot
        table[x=KiB_count,y=CTL_error_rate,col sep=comma] {ERR_1M.csv};
        \addplot
        table[x=KiB_count,y=SPR_error_rate,col sep=comma] {ERR_1M.csv}; 
        \addplot
        table[x=KiB_count,y=INT_error_rate,col sep=comma] {ERR_1M.csv};
        \addplot
        table[x=KiB_count,y=REB_error_rate,col sep=comma] {ERR_1M.csv};
      \end{axis}
    \end{tikzpicture}
\end{subfigure}
\hfill
\begin{subfigure}[b!]{0.31\textwidth}
    \begin{tikzpicture}
      \begin{axis}[
          width=\linewidth, 
          grid=major, 
          grid style={dashed,gray!30}, 
          xlabel=5MiB Download progress (KiB),
          ylabel=Failure rate (\%),
          no markers,
          every axis plot/.append style={very thick},
          scaled x ticks=false,
          scaled y ticks=false
        ]
        \addplot
        table[x=KiB_count,y=CTL_error_rate,col sep=comma] {ERR_5M.csv};
        \addplot
        table[x=KiB_count,y=SPR_error_rate,col sep=comma] {ERR_5M.csv}; 
        \addplot
        table[x=KiB_count,y=INT_error_rate,col sep=comma] {ERR_5M.csv};
        \addplot
        table[x=KiB_count,y=REB_error_rate,col sep=comma] {ERR_5M.csv};
      \end{axis}
    \end{tikzpicture}
\end{subfigure}

\hspace{0.0cm}
\vspace{-12pt}

\caption{{\bf Failure rate:} Percentage of attempted downloads that failed with standard Tor, REB, Spring, and Interspace 
}
\label{fig-err}
\end{figure*}

\subsection{Bandwidth overhead}

To visualize bandwidth overhead, we make a distinction between bytes of {\em content} received during a download and bytes of {\em padding} received. This allows us to examine the bytes of padding that have been received so far at any point in a download as a percentage of the bytes of content received. This is often referred to as {\em receive} bandwidth overhead; similarly, {\em send} bandwidth overhead is the padding sent as a percentage of the bytes of content sent, and total overhead takes both bytes sent and bytes received into account.

During each simulation, we measured the receive bandwidth overhead at various times throughout every download made by the benchmark clients. After filtering out the partial results of failed downloads, we calculated the median at each time for the three download sizes with standard Tor (the {\em control}), REB, Spring, and Interspace. We omit the results for REB since its median receive bandwidth overhead was 0\% for every download size. We also exclude all 50 KiB results as overhead was 0\% for the control and with each defense; the data for the 1 MiB and 5 MiB download sizes is shown in Figure~\ref{fig-rbw}.

For both download sizes, we observed lower values than those originally reported by the authors of the three defenses (for REB, 83\%~\cite{mathews2018understanding}; for Spring, 89\%; and for Interspace, 88\%~\cite{pulls2020towards}).  This is likely due to two factors: first, the different traffic patterns induced by single-file downloads as opposed to the web page downloads used as a basis for comparison in previous work will naturally lead to different patterns of adaptive cover traffic; and second, we observed a large fraction of failed downloads apparently caused by errors in the Tor circuit padding framework code, which we explore in the next section.

\subsection{Delay}

To measure delay, we recorded a timestamp and number of bytes each time data was received by a benchmark client. This allowed us to determine the time taken to reach any given byte count; that is, to examine the progress of each download over time. We aggregated these results over all simulations to obtain the median time to reach a number of progress points for the control and three defenses. In Figure~\ref{fig-ttb}, we compare the results for the 50 KiB, 1 MiB and 5 MiB download sizes.

It may be noted that the 50 KiB results are quite counterintuitive: the control actually had the highest median latency to every byte, and the median time to reach the 50 KiB mark for REB specifically was 5.5\% shorter than that of the control. The results for 5 MiB downloads are similar: although Interspace appears to have caused some additional delay, REB and Spring had shorter median times to 5 MiB than the control.

As with the bandwidth overhead data in Figure~\ref{fig-rbw}, we note that the time-to-byte data in Figure~\ref{fig-ttb} does not include the partial results of failed downloads. Since REB had a very high failure rate but little padding overhead, it is likely that download failures kept the total load on the network below that of the control throughout each simulation, allowing downloads that did succeed to complete more quickly.

Similarly, although Spring and Interspace involved more padding overhead, they both had very high failure rates for the 1 MiB and 5 MiB download sizes. As different benchmark clients performed downloads of all three sizes in parallel, it is likely that these download failures reduced the load on the network enough to affect the timing of successful downloads, including 50 KiB downloads.

%
%
To further illustrate this, we note that the time-to-byte data for 5 MiB downloads would seem to indicate that Spring had a median time to 5 MiB that was 9.6\% lower than that of the control. However, Spring had a 99\% failure rate for the 5 MiB download size, whereas the control had a 25\% failure rate, so this data is based on a much smaller pool of downloads that likely occurred when there was less load on the network.

\begin{wrapfigure}{r}{0.7\columnwidth}
    \begin{tikzpicture}
      \begin{axis}[
          width=\linewidth, 
          grid=major, 
          grid style={dashed,gray!30}, 
          xlabel=Padding count (cells),
          ylabel=Download time (ms),
          legend pos=north east,
          legend style={font=\tiny},
          every axis plot/.append style={very thick},
          scaled x ticks=false,
          scaled y ticks=false
        ]
        \addplot+[only marks,mark=+]
        table[x=INT_padding_count,y=INT_download_time,col sep=comma] {SCATTER_1M.csv};
        \addplot+[only marks,mark=x]
        table[x=SPR_padding_count,y=SPR_download_time,col sep=comma] {SCATTER_1M.csv};
        \legend{Interspace,Spring}
      \end{axis}
    \end{tikzpicture}
\vspace{-12pt}
\caption{Download time vs padding count (1 MiB)}
\label{fig-scatter1m}
\vspace{-12pt}
\end{wrapfigure}
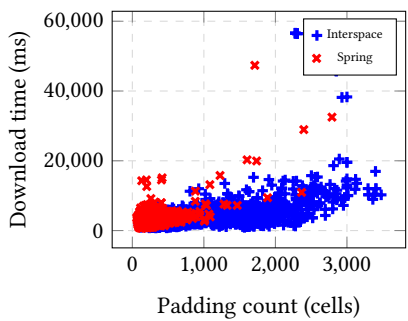
The 1 MiB download size data comes closer to reflecting our expectations: it indicates that Spring and Interspace did add delay, although REB remained similar to the control. At the 25\% mark, Spring had a median latency overhead of 2.7\% and Interspace of 6.3\%; at completion,  Spring and Interspace had median latency overheads of 1.6\% and 7.8\%, respectively. REB finished slightly below the control, but had the highest median failure rate (89\%), which supports our conclusions about download failures.  Figure~\ref{fig-scatter1m} also shows this trend, with download time increasing as a function of padding cells, but with an R$^2$ of only 0.21 for Interspace and 0.28 for Spring.

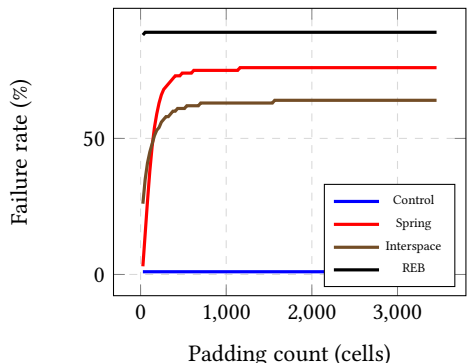
\begin{figure}
    \begin{tikzpicture}
      \begin{axis}[
          width=0.8\linewidth, 
          grid=major, 
          grid style={dashed,gray!30}, 
          xlabel=Padding count (cells),
          ylabel=Failure rate (\%),
          legend style={font=\tiny},
          legend pos=south east,
          no markers,
          every axis plot/.append style={very thick},
          scaled y ticks=false
        ]
        \addplot
        table[x=padding_count,y=CTL_error_rate,col sep=comma] {PAD_ERR_1M.csv};
        \addplot
        table[x=padding_count,y=SPR_error_rate,col sep=comma] {PAD_ERR_1M.csv};
        \addplot
        table[x=padding_count,y=INT_error_rate,col sep=comma] {PAD_ERR_1M.csv};
        \addplot
        table[x=padding_count,y=REB_error_rate,col sep=comma] {PAD_ERR_1M.csv};
        \legend{Control,Spring,Interspace,REB}
      \end{axis}
    \end{tikzpicture}
%


\vspace{-12pt}
\caption{Failure rate vs padding count (1 MiB)
}
\label{fig-pad-err}
\vspace{-12pt}
\end{figure}
Although the precise mechanisms by which these failures occurred are unclear, we found that downloads were more likely to fail as the total number of padding cells sent and received increased (except REB, since it had negligible padding overhead) as seen in Figure~\ref{fig-pad-err}. That is, most of the downloads which did succeed involved a relatively low number of padding cells. This means that, in the absence of failures, the delay incurred by each defense should be expected to be higher than what we report.

Ultimately, even though we can't conclude that these defenses would result in a similar failure rate on the live Tor network, we suspect that latency overhead would be greater than what we observed in practice. Both cases represent potentially significant impacts on the usability of the Tor network, suggesting that rigorous evaluation of all potential usability factors is necessary before the deployment of any defense.



%% file: 04-conclusion.tex

Due to the prevalence of download failures in our simulations, it is difficult to draw precise conclusions about the relationship between padding overhead and delay, but it is clear that padding-based defenses place additional stress on the Tor network, leading to nonzero added delays. This is especially true since download {\em failures} will either lead to higher delays as users reload pages and resources that fail to download, or simply lead to more severe user frustrations than those caused by the increased latency WF defense designers hope to avoid.

Our results highlight the fact that WF defenses require more complete evaluation than simple trace-based simulation or single-edge deployments, because these evaluations fail to capture the interaction between multiple defended circuits in the Tor network. They also suggest that the existing circuit padding framework may not be ready for deployment on the live network. Finally, in light of this need for further evaluation, it may be desirable to consider allowing cell delays, since adding padding already incurs latency, and cell delays may in fact reduce the resource stress caused by WF defenses.

Thus, an interesting avenue for future work is to explore how the circuit padding framework can be modified to allow delays, and to implement more recent low-overhead defenses such as FRONT~\cite{gong2020zero}, RegulaTor~\cite{holland2022regulator} and Surakav~\cite{gong2022surakav}.  This will also allow more complete evaluations of these defenses.
Similarly, considering our observation that TorNetTools measurements lead to different bandwidth overheads than previous workloads, another interesting avenue for future work is to explore the overhead of WF defenses on more realistic workloads.  Time-to-byte measurements can help to normalize comparisons across workloads, but other metrics based on time-to-event for various browser events may be useful as well. 
